\documentclass[runningheads]{llncs}

\usepackage[symbol]{footmisc}
\usepackage{multirow}

\usepackage{array}
\newcolumntype{L}[1]{>{\raggedright\let\newline\\\arraybackslash}m{#1}}
\newcolumntype{C}[1]{>{\centering\let\newline\\\arraybackslash}m{#1}}
\newcolumntype{R}[1]{>{\raggedleft\let\newline\\\arraybackslash}m{#1}}
\usepackage{pifont}
\usepackage{amsmath}
\usepackage{algorithm}
\usepackage{wrapfig,lipsum,booktabs}
\usepackage[noend]{algpseudocode}
\usepackage{float}
\usepackage[normalem]{ulem}
\usepackage{comment}
\usepackage{placeins}
\newcommand{\tablestyle}[2]{\setlength{\tabcolsep}{#1}\renewcommand{\arraystretch}{#2}\centering\footnotesize}

\newcommand{\app}{\raise.17ex\hbox{$\scriptstyle\sim$}}

\newcolumntype{x}[1]{>{\centering\arraybackslash}p{#1pt}}

\newlength\savewidth\newcommand\shline{\noalign{\global\savewidth\arrayrulewidth
  \global\arrayrulewidth 1pt}\hline\noalign{\global\arrayrulewidth\savewidth}}
\makeatletter\renewcommand\paragraph{\@startsection{paragraph}{4}{\z@}
  {.5em \@plus1ex \@minus.2ex}{-.5em}{\normalfont\normalsize\bfseries}}\makeatother

\usepackage{times,graphicx,amsmath,amssymb,booktabs,tabulary,multirow,overpic,xcolor}
\usepackage[caption=false]{subfig}
\usepackage[british,english,american]{babel}

\usepackage[font=small,labelfont=bf,tableposition=bottom]{caption}

\usepackage{graphicx}
\usepackage{graphicx}
\begin{document}
\title{Volumetric Attention for 3D Medical Image Segmentation and Detection}
%
%
\author{Xudong Wang\inst{1,2}\thanks{This work was fully conducted during the internship in 12 Sigma Technologies, USA.} \and
Shizhong Han\inst{1} \and
Yunqiang Chen\inst{1} \and \\ 
Dashan Gao\inst{1} \and
Nuno Vasconcelos\inst{2}}
\authorrunning{X. Wang et al.}
%
\institute{12 Sigma Technologies, San Diego, USA \and
Dept. of Electrical and Computer Engineering, Univ. of California, San Diego, USA 
\email{\{xuw080,nuno\}@ucsd.edu} \email{\{Shan,yunqiang,dgao\}@12sigma.ai}
}
\maketitle              
\begin{abstract}
A volumetric attention(VA) module for 3D medical image segmentation and detection is proposed. VA attention is inspired by recent advances in video processing, enables 2.5D networks to leverage context information along the z direction, and allows the use of pretrained 2D detection models when training data is limited, as is often the case for medical applications. Its integration in the Mask R-CNN is shown to enable state-of-the-art performance on the Liver Tumor Segmentation (LiTS) Challenge, outperforming the previous challenge winner by 3.9 points and achieving top performance on the LiTS leader board at the time of paper submission. Detection experiments on the DeepLesion dataset also show that the addition of VA to existing object detectors enables a 69.1 sensitivity at 0.5 false positive per image, outperforming the best published results by 6.6 points.

\keywords{Volumetric Attention \and 3D Images \and LiTS \and DeepLesion.}
\end{abstract}

\footnotetext[2]{Accepted by International Conference on Medical Image Computing and Computer-Assisted Intervention (MICCAI), 2019.}%

\section{Introduction}

A natural solution to 3D medical image segmentation and detection problems is to rely on 3D convolutional networks, such as the 3D U-Net of \cite{cciccek20163d} or the extended 2D U-Net of \cite{ronneberger2015u}. However, current GPU memory limitations prevent the processing of 3D volumes with high resolution. This is problematic, because the use of low-resolution volumes leads to low precision or miss-detection of small lesions and tumors and blur in lesion mask predictions, especially on boundaries. Hence, there is a need to trade-off the spatial resolution of each 2D slice for the number of slices processed. This implies a trade-off between the precision with which segmentation or detection can be performed and the amount of contextual information, in the z direction, that can be leveraged. A popular solution is to a use a 2D network to segment or detect the structures of interest in 2D or 2.5D slices and then concatenate the results to build a 3D segmentation mask or bounding box.

Christ et al. proposed a 2D U-Net for liver and tumor segmentation, followed by a conditional random field for segmentation refinement \cite{christ2016automatic}. Li et al. proposed a hybrid Dense 2D/3D UNet of three-stages \cite{li2018h}. They found that a pre-trained 2D model can significantly boost  performance of their network. Han proposed a 2.5D (adjacent slices) residual U-Net for liver lesion segmentation \cite{han2017automatic}. These approaches are limited by the lack of contextual information. Since even human experts need to inspect multiple slices to reach confident assessments of confusing lesions, this is likely to upper bound their performance.
To address this problem, Yan et al. \cite{yan20183d} proposed a 3D context enhanced region-based CNN. However, their method is based on a region proposal network (RPN) and cannot be implemented as a single-stage detector, such as SSD and YOLO, or a segmentation network, such as U-Net, without an RPN component. Furthermore, because only the feature map derived from a central image is processed by the RPN to generate proposals, the proposal generation process has no access to 3D context. Given that missed proposals can not be recovered, this places an upper bound on detection performance.

\begin{figure}[t!]
  \includegraphics[width=\linewidth]{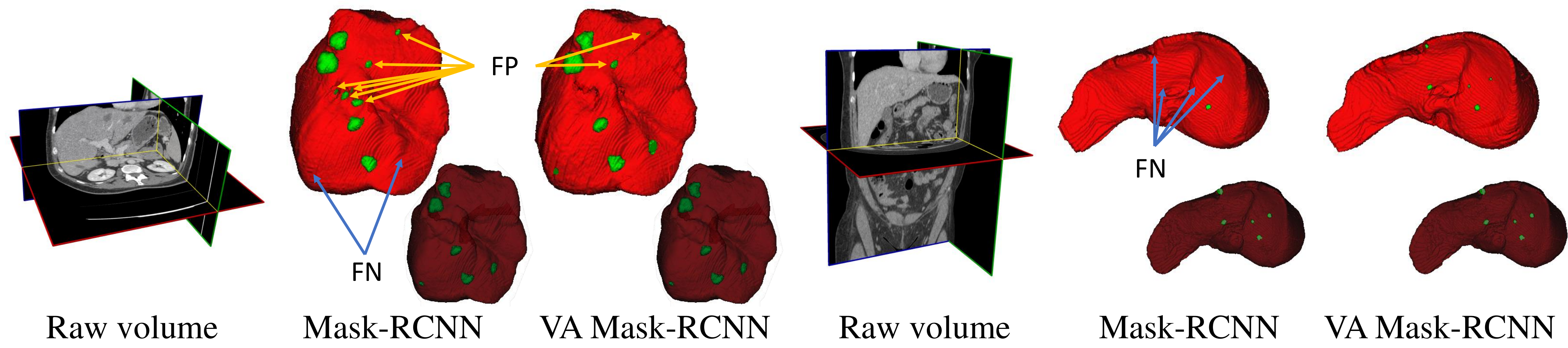}
  \caption{Comparison of 3D segmentations by the Mask-RCNN and the proposed VA Mask-RCNN on the  LiTS \texttt{val}  set. Red denotes  segmented  liver, green segmented lesions. 3D ground truth is shown on the bottom right, with liver in dark red and lesions in dark green. Left: while the Mask-RCNN misses two lesions (false nagative, FN) and has has six false positive (FP) instances, the VA Mask-RCNN detects all lesions with only two FPs. Right: VA Mask RCNN detects 5 very small lesions, 4 of which are missed by the Mask-RCNN. These examples illustrate how the VA module {\it both\/} enhances small lesion prediction and enables the network to avoid false positives. (best viewed in color)}
\label{fig:3d_gt_compare}\vspace{-5mm}
\end{figure}

In this work, we propose to address these limitations with ideas inspired by recent video processing work, where a similar problem is posed by the need to trade off the modeling of long-range dependencies between video frames and the  spatial resolution of each frame. The proposed approach is inspired by \cite{wang2018non}, which added a non-local network to a 3D convolutional network (C3D/I3D) for video classification, using a spacetime dependency/attention mechanism. We generalize this method into a flexible and computationally efficient Volumetric Attention (VA) module, which sequentially infers 3D enhanced attention maps along two separate dimensions, channel and spatial. The attention maps produced by this module are multiplied by the input feature map to enable adaptive feature refinement, using a 2D network. Similar to \cite{hu2017squeeze} and \cite{woo2018cbam}, global spatial pooling and global channel pooling are used to reduce computational cost. 

The VA module has several interesting properties. First, it enables the processing of high spatial resolution images, while leveraging contextual information over multiple slices of the 3D CT volume. Second, it can be combined with any CNN architecture, including one-stage and two-stage detectors and segmentation networks. Third, it is computationally efficient,  due to extensive use of spatial and channel pooling. Fourth, because the VA module can operate on image sub-regions, it can also benefit RPN networks. Fifth, since VA can be used with 2D networks, it can leverage pre-trained 2D CNN weights for transfer learning. 
The proposed VA attention module is implemented within the Mask-RCNN, leading to an architecture denoted the VA Mask-RCNN.
As illustrated in Fig.\ref{fig:3d_gt_compare}, this not only reduces segmentation false positives, but also enables the retrieval of very small lesions that are missed by the Mask-RCNN model. The VA Mask-RCNN is shown to obtain state-of-the-art performance, 74.1 dice per case, on the LiTS liver tumor segmentation challenge $\texttt{test}$ set, significantly outperforming (3.9 points) the winner of last year's challenge. It is the top method on the challenge leaderboard at the time of submission of this paper.  To assess the generalization ability of the VA Mask-RCNN to 3D CT volumes, we have also performed experiments on the DeepLesion dataset. The VA Mask-RCNN achieved a sensitivity of 77.22 at 1 false positives(FPs)/image, outperforming the best published results by $\sim$4 points.

\section{Volumetric Attention}
\begin{figure}[t!]
  \centering
  \includegraphics[width=0.85\linewidth]{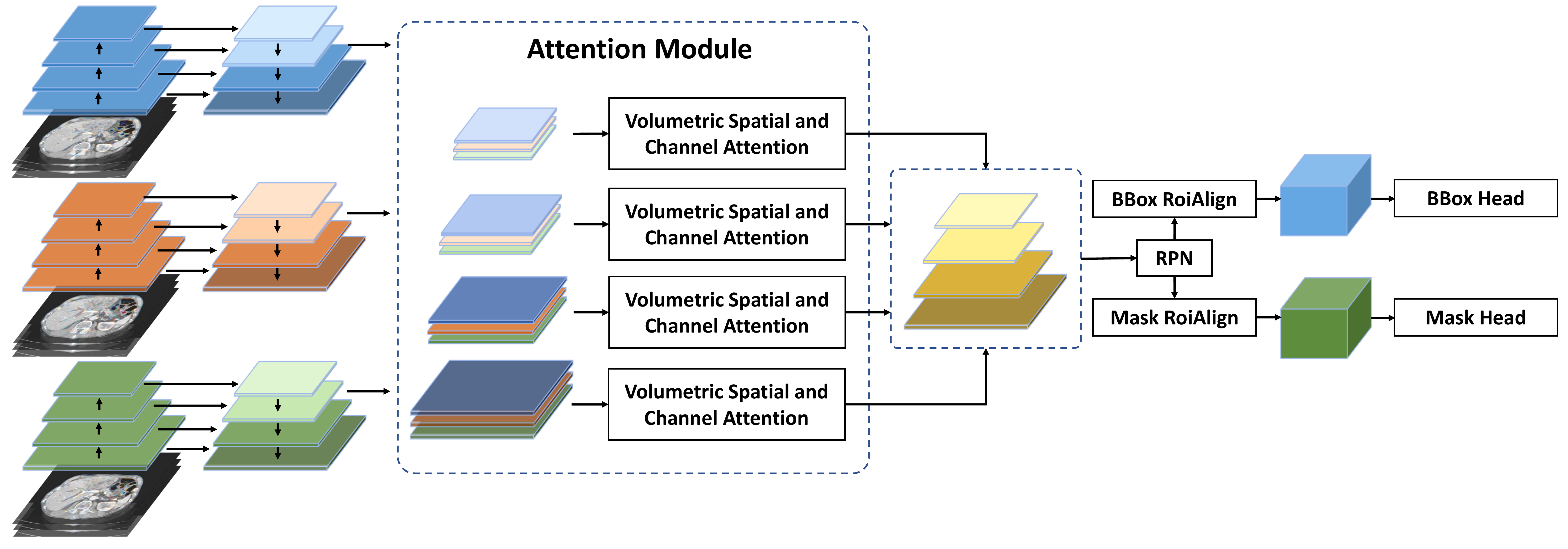}
  \caption{Architecture of the Volumetric Attention(VA) Mask-RCNN. Three continuous 2.5D images, each composed of 3 adjacent slices, are shown as example.}
\label{fig:CrossSliceChannelandSpatialAttention}\vspace{-3mm}
\end{figure}
\begin{figure}[t!]
  \centering
  \includegraphics[width=0.8\linewidth]{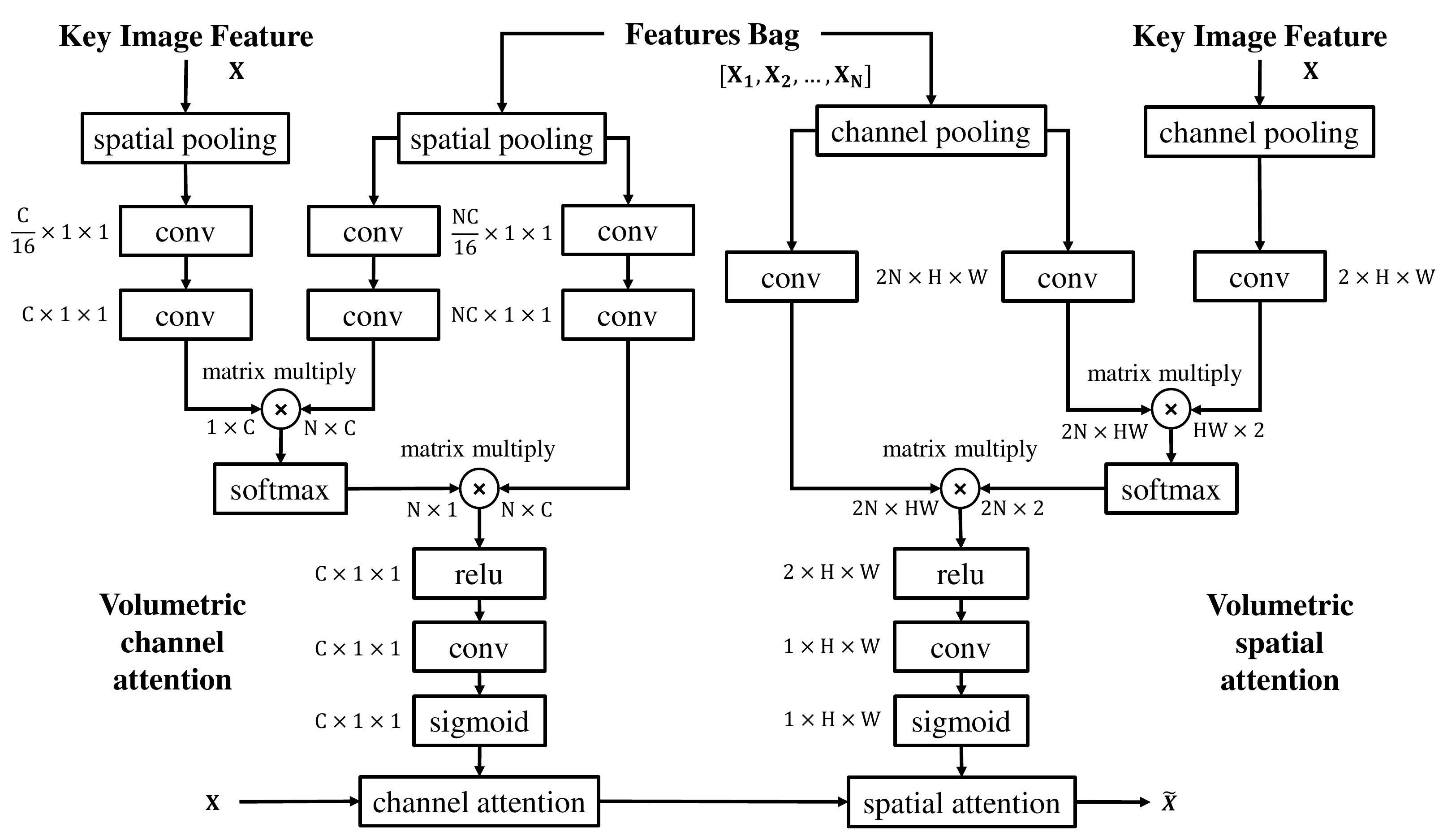}
  \caption{Volumetric Spatial and Channel Attention Module. N is the bag size, C, H, W the feature map channel size, height and width, respectively. Spatial and channel pooling are used to reduce computation.}
\label{fig:Scheme_CrossSliceChannelandSpatialAttention}\vspace{-3mm}
\end{figure}

The overall architecture of the VA Mask R-CNN is shown in Fig.\ref{fig:CrossSliceChannelandSpatialAttention}. The VA attention module operates on the Mask R-CNN feature pyramids extracted from a {\it target\/} 2.5D image, where detection takes place, and neighboring {\it contextual\/} 2.5D images. The 2.5D images are each composed of 3 adjacent slices. The attention module has three components: bag of long-range features, volumetric channel attention, and volumetric spatial attention. Unlike the self-attentive feature map of \cite{wang2018non}, VA uses long-range features from neighboring slices, which are combined with the feature map of the target slice to generate spatial and channel attention responses. A detailed scheme of the attention module is given in Fig. \ref{fig:Scheme_CrossSliceChannelandSpatialAttention}. We next discuss the three components combined with Mask-RCNN in detail.

\subsection{Bag of Long-range Features}
To account for dependencies along the z direction of the 3D CT volume, the VA Mask R-CNN complements the target 2.5D image, with neighboring images, both above and below the target image. These are denoted as contextual images.
The features extracted from these images are concatenated for each level of the spatial pyramid, according to
\begin{equation}
\textbf{X}_{long}^{i} = [\textbf{X}_{1},\textbf{X}_{2},...,\textbf{X}_{N}] \in \mathbb{R}^{N \times {C^{i}} \times {H^{i}} \times {W^{i}}},
\label{eq:func_1}
\end{equation}
where $i$ is the pyramid level, ${C^{i}} \times {H^{i}} \times {W^{i}}$ its dimensions (chanel, height, and width, respectively), $\textbf{X}_{long}^{i}$ the corresponding bag of long-range features, and $N$ the number of contextual images. The features $\textbf{X}_{k}$ are sorted by the order of the corresponding images along the z direction of the 3D volume. 

\subsection{Volumetric Channel Attention}
\label{sec:cross_channel}
This attention mechanism is inspired by that of \cite{hu2017squeeze,wang2018non}. The bag of features $\textbf{X}_{long} \in \mathbb{R}^{N \times {C} \times {H} \times {W}}$ and corresponding target image feature map $\textbf{X}_{tgt} \in \mathbb{R}^{{C} \times {H} \times {W}}$ are each subject to a global average pooling operator $\textbf{F}_{avg}^{c}$. Following \cite{hu2017squeeze}, computation is reduced by replacing the linear embedding layer of the original non-local blocks of~\cite{wang2018non} by two $1 \times 1$ convolutional layers with reduction ratio of 16. This is implemented as $\textbf{F}_{emb}^{c}(\mathbf{X}) = W_2 \delta(W_{1}\textbf{F}_{avg}^{c}(\mathbf{X}))$, where $W_{1} \in \mathbb{R}^{\frac{C}{16} \times {C}}$, $W_{2} \in \mathbb{R}^{C \times \frac{C}{16}}$ and $\delta$ is the RELU function.
The slices attention signal is finally computed with a softmax  
\begin{equation} 
\textbf{S}_{att}^{c} = \text{softmax}(\textbf{F}_{emb}^{c}(\mathbf{X}_{tgt}) \cdot \textbf{F}_{emb}^{c}(\mathbf{X}_{long})) \in \mathbb{R}^{1 \times {N}}
\end{equation}
along dimension $N$, where $\textbf{F}_{emb}^{c}(\mathbf{X}_{tgt}) \in \mathbb{R}^{1 \times {C}}$, $\textbf{F}_{emb}^{c}(\mathbf{X}_{long}) \in \mathbb{R}^{C \times N}$ and $\cdot$ refers to matrix multiplication. The slice attention signal $\textbf{S}_{att}^{c}$ is then applied to $\textbf{F}_{emb}^{c}(\mathbf{X}_{long}) \in \mathbb{R}^{N \times {C}}$ according to
$\textbf{S}_{att}^{c} \cdot \textbf{F}_{emb}^{c}(\mathbf{X}_{long})$ and this is followed by a relu layer, a $1 \times 1$ conv layer and a sigmoid layer, to learn a nonlinear interaction $\mathbf{S}_{c} \in \mathbb{R}^{C \times {1} \times {1}}$ between channels. Then channel-wise multiplication is applied on $\mathbf{X}_{tgt} \in \mathbb{R}^{{C} \times {H} \times {W}}$.


\subsection{Volumetric Spatial Attention}
The volumetric spatial attention module uses max and average pooling to shrink feature maps along the channel dimension,  concatenating them into two channel feature maps $\textbf{F}_{pool}^{s}(\mathbf{X}) = [\textbf{F}_{max}^{s}(\mathbf{X}), \textbf{F}_{avg}^{s}(\mathbf{X})] \in \mathbb{R}^{{2} \times {H} \times {W}}$. An embedding function is then implemented as $\textbf{F}_{emb}^{s}(\mathbf{X})=W\textbf{F}_{pool}^{s}(\mathbf{X})$, where $W$ is a learned convolutional weight layer. The slice attention signal is finally computed with a softmax
\begin{equation} 
\textbf{S}_{att}^{s} = \text{softmax}(\textbf{F}_{emb}^{s}(\mathbf{X}_{tgt}) \cdot \textbf{F}_{emb}^{s}(\mathbf{X}_{long})) \in \mathbb{R}^{1 \times {N}}
\end{equation}
along dimension $N$. A spatial attention map $S_{s} \in \mathbb{R}^{ 1 \times {H} \times {W}}$ is then generated with an architecture similar to that of Section \ref{sec:cross_channel} and element-wise multiplied with $\mathbf{X}_{tgt} \in \mathbb{R}^{{C} \times {H} \times {W}}$.

\section{Experiments}
\label{sec:exp}
The volumetric attention was evaluated on two public datasets, Liver Tumor Segmentation (LiTS)~\cite{bilic2019liver} and DeepLesion\cite{yan2018deeplesion}. All experiments used a PyTorch implementation \cite{mmdetection2018} of the Mask-RCNN and Faster R-CNN. Unless otherwise noted, all  hyperparameters are as in \cite{lin2017feature} for the Faster-RCNN and \cite{he2017mask} for the Mask-RCNN.

\begin{figure}[t!]
  \centering
  \includegraphics[width=0.95\linewidth]{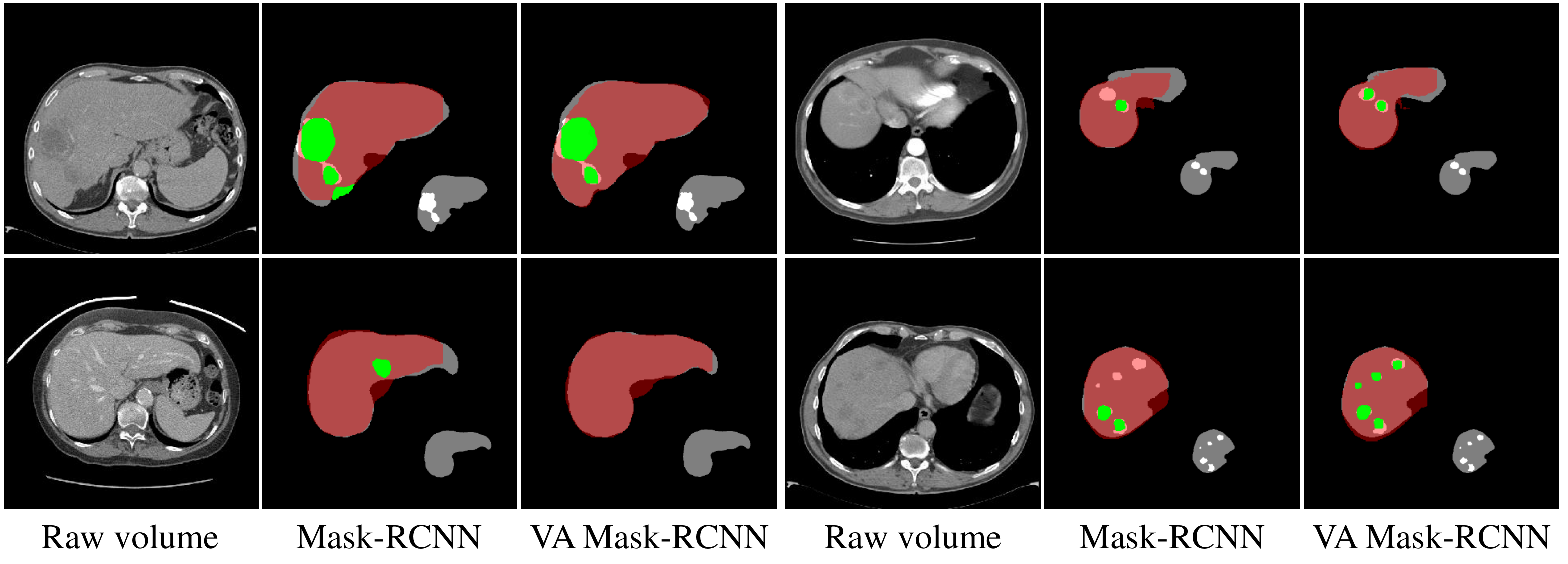}
  \caption{2D visualization of segmentations by Mask-RCNN and VA Mask R-CNN on  LiTS \texttt{val}  set. Segmented  liver is shown in red and lesions in green. Zoomed out ground truth masks are shown on bottom right, with liver in gray and lesions in white. The VA Mask-RCNN produces smoother segmentation boundaries and lower FP and miss rates. In the top left, the gallbladder area is easily confused with the lesion area. VA Mask-RCNN leverages contextual slices to  remove this FP. (best viewed in color and zoom in for details)}
\label{fig:gt_compare}\vspace{-3mm}
\end{figure}

\subsection{Datasets and Evaluation}
LiTS is a dataset of liver lesions, including 131 training and 70 test CT scans, acquired in six different clinical sites using different protocols and scanners. Lesion segmentation performance is evaluated and ranked by the Dice coefficient per volume, averaged over all test cases. For additional insight on the quality of the segmentation, we also break down the average Dice/lesion per lesion size: the coefficients measured for small (diameter $<$ 15 mm), medium (diameter between [15mm, 30mm] and large (diameter $>$ 30mm) legions are denoted as $\text{Dice}_{s}$, $\text{Dice}_{m}$ and $\text{Dice}_{l}$ respectively. 
DeepLesion is a dataset with a larger variety of lesions, including 33,688 bookmarked radiology images from 10,825 studies of 4,477 unique patients. For each bookmarked image, a bounding box is generated to indicate the location of each lesion. We use the official split (70$\%$ training, 15$\%$ validation and 15$\%$ test) at the patient level, for training and testing. For consistency with prior art, detection results are evaluated with the False Positives (FPs) per Image metric

\subsection{LiTS Experiments}
\noindent{\bf Pre-processing:} For 3D liver/lesion detection and segmentation, we stack three adjacent axial slices into a 3-channel image and apply the Mask-RCNN to detect and segment the liver/lesion for the center slice. 3D segmentation results are then obtained by stacking the masks generated for all slices. The Mask-RCNN is trained to detect both liver and lesions, to enable the removal of false lesions outside the liver by simply computing the logical AND of the predicted liver and lesion masks. Since the focus of this task is on the liver and lesions, the CT scan's Hounsfield unit (HU) is clamped between [-200, 300] and normalized to a floating point between [0, 1]. Each slice is scaled to 1024{$\times 1024$} pixels and the slice-thickness resampled to 1.5mm. 

\noindent{\bf Benchmark results:} To evaluate performance on LiTS, the feature bag size of~(\ref{eq:func_1}) was set to 9, the weights of the feature extractor and RPN copied from detectors pre-trained 
\begin{wraptable}{r}{6cm}\vspace{-8mm}
\tablestyle{2.0pt}{1.2}
\scriptsize
\begin{tabular}{l|l|x{40}}
Team & Model & Dice per case \\ [.1em]
\shline
3D U-Net(Ours) \cite{cciccek20163d} & 3D U-Net & 55.0 \\
G. Chlebus \cite{chlebus2017neural} & 2D U-Net & 65.0 \\
E. Vorontsov et al. \cite{vorontsov2018liver} & 2D + 3D FCN & 65.0 \\
Y. Yuan \cite{yuan2017hierarchical} & Deconv-Conv Net & 65.7 \\
X. Han \cite{han2017automatic} & 2D U-Net & 67.0 \\
LeHealth & - & 70.2 \\
Mask-RCNN(Ours)\cite{he2017mask} & Mask-RCNN & 70.3 \\
X. Li et al.\cite{li2018h} & H-DenseUNet & 72.2 \\
\hline
VolumetricAttention & VA Mask-RCNN & \bf{74.1} \\
\end{tabular}\vspace{-2mm}
\caption{Comparison with LiTS Challenge leaderboard, as of July 1st, 2019}
\label{table:lits_test}\vspace{-8mm}
\end{wraptable}
on the MS-COCO and DeepLesion datasets, and the smallest image scale set to 1024. Table \ref{table:lits_test} presents a copy of the LiTS leaderboard, at the time of submission of this paper. All algorithms are evaluated on the LiTS \texttt{test} set. The VA Mask R-CNN achieves state-of-the-art performance, with 74.10 dice per case. This outperforms the previous LiTS challenge winner by 3.9 points and the best published results by 1.9 points. 

\noindent{\bf Ablation study and evaluation:} To better understand the proposed architecture, the LiTS dataset was split, using 75{\%} of the \texttt{train} data to create a training set and the remaining 25{\%} as a \texttt{val} set for a local ablation study.  Table \ref{table:lits_val} summarizes the resulting dice per volume, averaged over all cases, and dice per cases, averaged over small, medium and large lesions. All these are control experiments, all hyper-parameters and settings remaining the same as in the benchmark experiments, unless otherwise noted. 

\noindent{\bf Benefits of VA attention:} Three conclusions can be drawn from Table \ref{table:diff_methods}. First, the 2D approaches outperform the 3D U-Net, even before addition of the VA attention module. This shows that 2D networks are at least competitive for 3D mask segmentation. Since the Mask-RCNN achieved the best performance on these experiments, we use it as base model in the remainder of the paper. It should, however, be pointed out that VA could equally be combined with the 2D U-Net. Second, the addition of the VA module further increases performance, increasing the Dice coefficient per case by 4.7 points. Third, this gain is especially large for small and medium lesions, e.g. 8 points for small lesions. Note how the lack of contextual information along the z direction severely compromises the small lesion performance of the mask R-CNN. Fig.\ref{fig:gt_compare} illustrates how VA attention enables the Mask R-CNN to reject confusing FP lesions and produce smoother segment boundaries.

\begin{table*}[t]
\footnotesize
\subfloat[Dice comparison.\label{table:diff_methods}]{
\tablestyle{1.pt}{1.05}\scriptsize\begin{tabular}{l|x{18}x{18}x{20}x{18}} & {Dice} & {$\text{Dice}_{s}$} & {$\text{Dice}_{m}$} & {$\text{Dice}_{l}$} \\
\shline
3D U-Net& 35.3 & 17.0 & 39.2 & 61.3 \\
2D U-Net& 48.8 & 39.7 & 58.2 & 68.3 \\
Mask-RCNN & 56.1 & 44.3 & 70.6 & 78.4 \\
\hline
Ours & \bf{60.8} & \bf{52.2} & \bf{71.4} & \bf{78.7} \\
\end{tabular}}
\subfloat[Pre-training dataset. \label{table:pretrain}]{
\tablestyle{1.pt}{1.05}\scriptsize\begin{tabular}{l|x{22}x{22}x{22}} & \multicolumn{3}{c}{\textit{Pre-training dataset}}\\
\shline
+ImageNet & \checkmark & \checkmark & \checkmark \\
+MS-COCO &  & \checkmark & \checkmark \\
+DeepLesion &  &  & \checkmark \\
\hline
\textbf{\textit{Dice per case}} & 60.8 & 61.9 & \bf{63.3}\\
\end{tabular}}
\subfloat[Influence of image scales.\label{table:scale}]{
\tablestyle{1.pt}{1.05}\scriptsize\begin{tabular}{x{25}|x{18}x{18}x{20}x{18}}
Scale & {$\text{Dice}$} & {$\text{Dice}_{s}$} & {$\text{Dice}_{m}$} & {$\text{Dice}_{l}$} \\
\shline
512 & 50.2 & 35.8 & 65.1 & 77.9 \\
800 & 61.1 & 52.1 & 71.6 & 79.3\\
1024 & \bf{63.3} & \bf{54.3} & \bf{73.7} & \bf{80.3} \\
1333 & \bf{63.5} & \bf{54.8} & \bf{73.5} & \bf{80.4} \\
\end{tabular}}\hspace{0.2mm}
\subfloat[Influence of VA modules.\label{table:each_module}]{
\tablestyle{1pt}{1.05}\scriptsize \begin{tabular}{l|x{18}x{18}x{20}x{18}} & {Dice} & {$\text{Dice}_{s}$} & {$\text{Dice}_{m}$} & {$\text{Dice}_{l}$} \\
\shline
Baseline & 56.1 & 44.3 & 70.6 & 78.4\\
\hline
+channel att & 61.5 & 52.2 & 72.7 & 78.7\\
+spatial att & \bf{63.3} & \bf{54.3} & \bf{73.7} & \bf{80.3}\\
\end{tabular}}
\subfloat[Influence of VA location. \label{table:possition}]{
\raggedright
\tablestyle{1pt}{1.05}\scriptsize\begin{tabular}{l|x{18}x{18}x{20}x{18}} & {Dice} & {$\text{Dice}_{s}$} & {$\text{Dice}_{m}$} & {$\text{Dice}_{l}$} \\
\shline
Baseline & 56.1 & 44.3 & 70.6 & 78.4\\
\hline
RPN & \bf{63.3} & \bf{54.3} & \bf{73.7} & \bf{80.3}\\
RCNN & 61.3 & 51.7 & 71.8 & 79.9\\
\end{tabular}}
\subfloat[Influence of number of slices. \label{table:slice_num}]{
\raggedright
\tablestyle{1pt}{1.05}\scriptsize\begin{tabular}{l|x{18}x{18}x{20}x{18}} 
\# Slices & {$\text{Dice}$} & {$\text{Dice}_{s}$} & {$\text{Dice}_{m}$} & {$\text{Dice}_{l}$} \\
\shline
$9(3\times3)$  & 61.7 & 52.2 & 71.6 & 79.5\\
$21(3\times7)$  & 62.5 & 52.6 & 72.2 & 79.8\\
$27(3\times9)$  & \bf{63.3} & \bf{54.3} & \bf{73.7} & 80.3\\
$33(3\times11)$  & 63.1 & 53.6 & 73.4 & \bf{80.6}\\
\end{tabular}}\vspace{1mm}
\caption{Evaluation on LiTS \texttt{val} set, in terms of dice per volume, averaged over all cases, and dice per lesions, averaged over small, medium and large lesions.}
\vspace{-6mm}
\label{table:lits_val}\vspace{-3mm}
\end{table*}

\noindent{\bf Influence of pre-training:} \cite{he2018rethinking} claims that ImageNet pre-training does not improve accuracy of networks trained with as few as 10k COCO images. As shown in Table~\ref{table:pretrain}, this does not hold for medical imaging where, due to the difficulties of collecting and labeling datasets, few datasets have 10k examples. Furthermore, while MS-COCO has $\sim$5 objects/image, this number is much smaller for medical image datasets. For LiTS the number is smaller than 1, especially when the 3D volume is split into 2D slices and these are considered different examples. Table~\ref{table:pretrain} shows that, in this case, ImageNet pre-training still has an important role in combating overfitting. Adding MS-COCO to the pre-training dataset further improves performance by 1.1 points. This is mostly because the  COCO tasks encourage the network to more accurately localize objects. Finally, due to the non-trivial domain shift between MS-COCO and LiTS, further pre-training on DeepLesion improves performance by an additional 1.4 points.

\noindent{\bf Image scales.} Table~\ref{table:scale} shows that larger image scales lead to better performance, especially for small lesions. However, performance saturates at a scale of 1333 pixels. This is only marginally better than a scale of 1024 but requires substantially more memory. For this reason, a scale of 1024 is adopted in the remainder of the paper.

\noindent{\bf Spatial vs. Channel Attention:} to compare the relative importance of the two attention mechanisms, the two modules were incrementally added to the 2D Mask-RCNN, with the results of Table \ref{table:each_module}. These experiments use 9 slices. The addition of channel attention enhances performance by more than 4 points, and the subsequent addition of spatial attention increases performance by another 2 points. In summary, both attention mechanisms are important.

\noindent{\bf Location of attention module:} the VA module can be added as shown in Fig.\ref{fig:CrossSliceChannelandSpatialAttention}, i.e. to the last stage of feature extraction, before the RPN, or after the bounding box ROI align and mask ROI align steps, i.e. before the RCNN. Table \ref{table:possition} shows that attention is more effective if introduced before the RPN. While this improves performance by 5.1 Dice points per case, the gain is only 1.7 points when attention is introduced after the RCNN. This shows that 3D context is important for high quality proposal generation. Since only RPN detected ROIs are used to crop feature maps, addition of attention after the RPN only improves the ability to reject FPs. In this case, attention cannot improve the retrieval of lesions that are otherwise missed.

\noindent{\bf Feature bags size:} Table \ref{table:slice_num} compares the network performance as the feature bag size 
varies between 3, 7, 9 and 11 images. While dice per case increases with feature bag size, the small and medium lesion performance starts to worsen beyond a bag size of 9. We thus adopt this size in the remaining experiments. We note, however, that for applications sensitive to inference time, smaller bag size may be preferable.
\vspace{-2pt}
\subsection{Extension Experiments on DeepLesion}


\begin{minipage}[t]{\textwidth}
\begin{minipage}[b]{0.49\textwidth}
\centering
\tablestyle{2.0pt}{1.2}
\scriptsize
\begin{tabular}{l|x{28}|x{12}x{12}x{12}}
Model & Backbone & 0.5 & 1 & 2\\ [.1em]
\shline 
Faster-RCNN\cite{girshick2015fast} & VGG-16 & 56.9 & 67.3 & 75.6\\
R-FCN\cite{dai2016r} & VGG-16 & 55.7 & 67.3 & 75.4\\
Improved R-FCN \cite{dai2016r} & VGG-16 & 56.5 & 67.7 & 76.9 \\
Data-level fusion, 11 slices & VGG-16 & 58.5 & 70.0 & 77.9\\
3-DCE,9 Slices\cite{yan20183d} & VGG-16 & 59.3 & 70.7 & 79.1\\
3-DCE,27 Slices\cite{yan20183d} & VGG-16 & 62.5 & 73.4 & 80.7\\
\hline
Faster-RCNN+VA, 9 Slices & ResNet50 & \bf{67.6} & \bf{75.6} & \bf{82.5}\\
Deformable Faster-RCNN+VA & ResNet50 & \bf{69.1} & \bf{77.9} & \bf{83.8}\\
\end{tabular}
\captionof{table}{Sensitivity($\%$) at 0.5, 1 and 2 FPs/image on the DeepLesion \texttt{test} set.}
\label{table:deeplesion_test}
\end{minipage}
    \begin{minipage}[b]{0.49\textwidth}
    \centering
    \tablestyle{2.0pt}{1.2}
    \scriptsize
    \begin{tabular}{l|x{30}|x{16}x{16}}
    Model & Backbone & 1 FPs  & AP$_{50}$ \\ [.1em]
    \shline 
    Faster-RCNN\cite{girshick2015fast} & ResNet152 & 77.4 & 64.9\\
    Faster-RCNN\cite{girshick2015fast} & ResNet101 & 75.1 & 61.8\\
    \hline
    Faster-RCNN\cite{girshick2015fast} & ResNet50 & 73.4 & 60.0\\
    Deformable Faster-RCNN\cite{dai17dcn} & ResNet50 & 76.3 & 62.4 \\
    \hline
    Faster-RCNN+VA & ResNet50 & 75.6 & 63.0\\
    Deformable Faster-RCNN+VCA & ResNet50 & 76.8 & 63.8\\
    Deformable Faster-RCNN+VSA & ResNet50 & 76.9 & 64.1\\
    Deformable Faster-RCNN+VA & ResNet50 & \bf{77.9} & \bf{65.0}\\
    \end{tabular}
    \captionof{table}{Sensitivity ($\%$) at 1 FPs/image and AP$_{50}$ on the DeepLesion \texttt{test} set.}
    \label{table:ablation}
    \end{minipage}
\end{minipage}\hspace{1mm}
To test the effectiveness of volumetric attention for the processing of 3D CT volume datasets, we performed some extension experiments on DeepLesion. This dataset enables the use of part of the 3D CT volume as context for 2D bounding box prediction on target slices. Since DeepLesion does not provide mask groundtruth, the VA module was implemented on Faster-RCNN-FPN and Deformable Faster-RCNN-FPN detectors, with ResNet50 backbones. As usual for DeepLesion, performance is evaluated with FPs per image. AP$_{50}$ is also presented in Table\ref{table:ablation}. All experiments in this section are based on training with the DeepLesion \texttt{train} and \texttt{val} sets, and testing on \texttt{test} set. Each 2.5D image is formed by concatenating 3 contiguous slices and scaled to 512$\times$ 512 pixels as in \cite{yan2018deeplesion}, the Faster-RCNN-FPN backbone is pretrained on ImageNet. Feature bag size is fixed to be 9, i.e. 3 continuous 2.5D images.

Table \ref{table:deeplesion_test} and Table \ref{table:ablation}, compare the proposed networks to several methods from the literature. The proposed networks achieve state of the art results, increasing sensitivity by 6.6 points at 0.5 FPs/image, 4.4 points at 1 Fp/image and 3.1 at 2 FPs/image. Table \ref{table:ablation}, shows that the proposed network with the ResNet50 backbone is comparable with the heavier Faster-RCNN with ResNet101 backbone. Independently adding Volumetric Channel Attention(VCA) and Volumetric Spatial Attention(VSA) to Deformable Faster-RCNN with ResNet50 can get 1.4 and 1.6 points performance increase separately, integrating VSA and VSA got 2.6 points performance improvement, this result is even slightly higher than much heavier Faster-RCNN with ResNet152 backbone.

\vspace{-2mm}
\section{Conclusion}
In this paper, we proposed a volumetric attention module that enables 2.5D methods to leverage contextual information along the z direction and the use of pretrained 2D detection models when training data is limited, as is often the case for medical applications. VA can be combined with any CNN architecture, including one-stage and two-stage detectors and segmentation networks. It was shown that 2.5D networks with VA achieve state of the art results for {\it both\/} lesion segmentation and detection.

%
%
%
%

\vspace{-1.5mm}
\bibliographystyle{splncs04}
\bibliography{egbib}





\end{document}